# Observation of the Wigner-Huntington Transition to Solid Metallic Hydrogen

Ranga Dias and Isaac F. Silvera

Lyman Laboratory of Physics, Harvard University, Cambridge MA 02138

We have studied solid hydrogen under pressure at low temperatures. With increasing pressure we observe changes in the sample, going from transparent, to black, to a reflective metal, the latter studied at a pressure of 495 GPa. We have measured the reflectance as a function of wavelength in the visible spectrum finding values as high as 0.90 from the metallic hydrogen. We have fit the reflectance using a Drude free electron model to determine the plasma frequency of 30.1 eV at T= 5.5 K, with a corresponding electron carrier density of $6.7 \times 10^{23}$ particles/cm$^3$, consistent with theoretical estimates. The properties are those of a metal. Solid metallic hydrogen has been produced in the laboratory.

Controlled nuclear fusion, production of metallic hydrogen, and high temperature superconductivity have been listed as the top three key problems of physics (*1*). These problems all involve hydrogen and its isotopes. The transition to solid metallic hydrogen (**SMH**) was envisioned by Wigner and Huntington (**WH**) over 80 years ago (*2*). They predicted that if solid molecular hydrogen were compressed to a sufficiently high density, there would be a first-order dissociative transition to an atomic lattice. Solid atomic hydrogen would be a metal with one electron per atom so that the conduction band would be half filled. Although WH's density for the transition was approximately correct, their predicted pressure of 25 GPa (100 GPa=1megabar) was way off. Today, modern quantum Monte-Carlo methods, as well as density functional theory (**DFT**), predict a pressure of ~400 to 500 GPa for the transition (*3-5*). The most likely space group for the atomic lattice is I4$_1$/amd (*5, 6*). Metallic hydrogen has been predicted to be a high temperature superconductor, first by Ashcroft (*7*), with critical temperatures possibly higher than room temperature (*8, 9*). Moreover, SMH is predicted to be metastable so that it may exist at room temperature when the pressure is released (*10*). If so, and superconducting, it could have an important impact on mankind's energy problems and would revolutionize rocketry as a powerful rocket propellant (*11*). In this paper we report the production of solid metallic hydrogen in the laboratory.

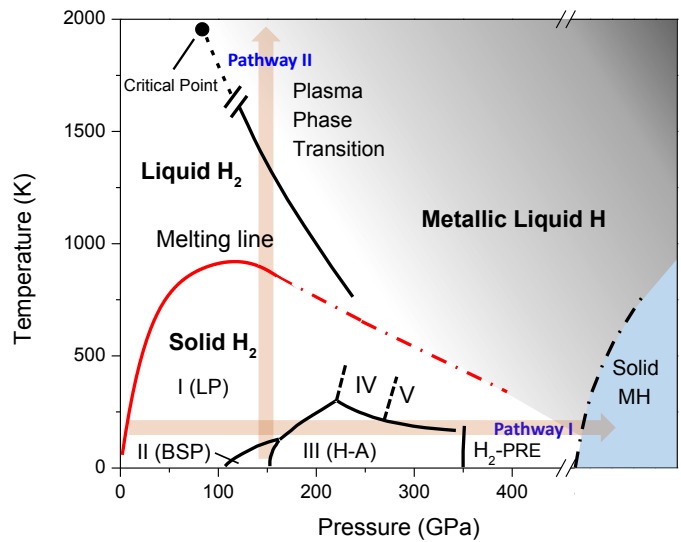

Fig. 1. A recent experimental/theoretical P-T phase diagram of hydrogen showing two pathways metallic hydrogen.

Wigner and Huntington predicted a simple phase diagram (**PD**); over the past decades the PD of metallic hydrogen (**MH**), shown in Fig. 1, has undergone enormous experimental and theoretical development as the pressure was increased ever higher. There are two pathways to MH shown in the figure. The WH transition to SMH follows pathway I and is the focus of this paper. Pathway II leads to a phase transition to liquid atomic metallic hydrogen, (also called the plasma phase transition) recently observed at static pressures in a diamond anvil cell (DAC) (*12*), further discussed in the Supplementary Materials (**SM**).

Pathway I transitions through a number of phases, not envisioned by WH. The low-pressure properties of solid molecular hydrogen are fascinating and many aspects have been reviewed elsewhere (*13*). In the low-pressure phase I, molecules are in spherically symmetric quantum states and form a hexagonal close packed (hcp) structure. Phase II, III, and IV are phases with structural changes and orientational order of the molecules. In addition it has been predicted that molecular hydrogen may become a metal at high pressures (*14*) (not shown in Fig. 1). A new phase in hydrogen has recently been observed at liquid helium temperatures called $H_2$-PRE (*15*) (also named VI at higher temperatures (*16*)), as it was believed to precede the transition to SMH. This phase transition was detected by changes in the infrared spectra. The highest reported pressure achieved in that experiment was 420 GPa, a pressure at which the sample was black in the visible; this is expected due to closing of its electronic band gap. A weak IR transmission signal was observed, meaning that it was not metallic. At those pressures the IR signal was too weak to measure spectra in our system. Pressure was determined from the Raman scattering shift of the diamond phonon in the stressed culet region of the diamonds, calibrated at high pressure by Akahama and Kawamura (**AK**) (*17, 18*). In that paper we used the AK 2010 calibration. For reference, we name that experiment MH-run-1.

In the current experiment, MH-run-2, we have carried out a rigorous strategy to achieve the higher pressures needed to transform to SMH. The principal limitation for achieving the required pressures to observe SMH in a DAC has been failure of the diamonds. Hydrogen is very diffusive at room temperature or higher and it can disperse into the confining gasket or the diamonds (at high pressure); diamonds become embrittled and fail. Diffusion is an activated process and is suppressed at low temperatures. In both MH experimental runs, the sample was maintained at liquid nitrogen or liquid helium temperatures. Diamond anvils used in DACs are generally polished with fine diamond powder on a polishing wheel. We believe that failure of diamonds can arise from microscopic surface defects created in the polishing process. We used type IIac conic synthetic diamonds (supplied by Almax Easy-Lab) with ~30 micron diameter culet flats. About 5 microns were etched off of the diamond culets using the technique of reactive ion etching, to remove defects from the surface. The diamonds were then vacuum annealed at high temperature to remove residual stress. Alumina is known to act as a diffusion barrier against hydrogen. The diamonds, with the mounted rhenium gasket, were coated with a 50 nm thick layer of amorphous alumina by the process of atomic layer deposition. We have had extensive experience with alumina coatings at high pressures and find that it does not affect or contaminate the sample, even at temperatures as high as ~2000 K (*12*). Finally, it is known that focused laser beams on samples at high pressures in DACs can lead to failure of the diamonds when the diamonds are highly stressed, even at laser powers as low as 10 mW. This is especially true if the laser light is in the blue region of the spectrum and is possibly due to laser induced growth of defects (*19*); another possibility is thermal shock to the stressed culet region, resulting from inadvertent laser heating. Moreover, a sufficiently intense laser beam, even at infra red (**IR**) wavelengths, can graphitize the surface of diamond. Thus, we study the sample mainly with very low power incoherent IR radiation from a thermal source, and minimize illumination of the sample with lasers when the sample is at very high pressures.

The goal for this experiment was to go to higher pressures than in our previous run that ended at ~420 GPa (*15*). The determination of pressure in the multi-megabar regime is challenging (see the SM). The sample was cryogenically loaded at 15 K and included a grain of ruby for pressure determination. The pressure was initially determined to ~88 GPa by ruby fluorescence using the scale of Chijioke et al (*20*); the exciting laser power was limited to a few mW. At higher pressures we measured the IR vibron absorption peaks of hydrogen with a Fourier transform infrared spectrometer with a thermal IR source, using the known pressure dependence of the IR vibron peaks for pressure determination (see SM). This was done to a pressure of ~335 GPa, while the sample was still transparent (see Fig. 2a). For higher pressures we used an alternate technique that did not require using a laser. Our DACs are equipped with strain gauges that measure the load on the sample. The load, or pressure, is increased by rotating a screw attached to the DAC in the cryostat by a long stainless steel tube, accessible at room temperature. We have previously studied the behavior of our DACs and found that the pressure is proportional to the rotation of the screw (see SM). With the next rotations of the screw the sample transitioned to the phase $H_2$-PRE and started turning black (see Fig. 2b). This darkening was also observed in MH-run-1, meaning that the pressure was ~400 GPa.

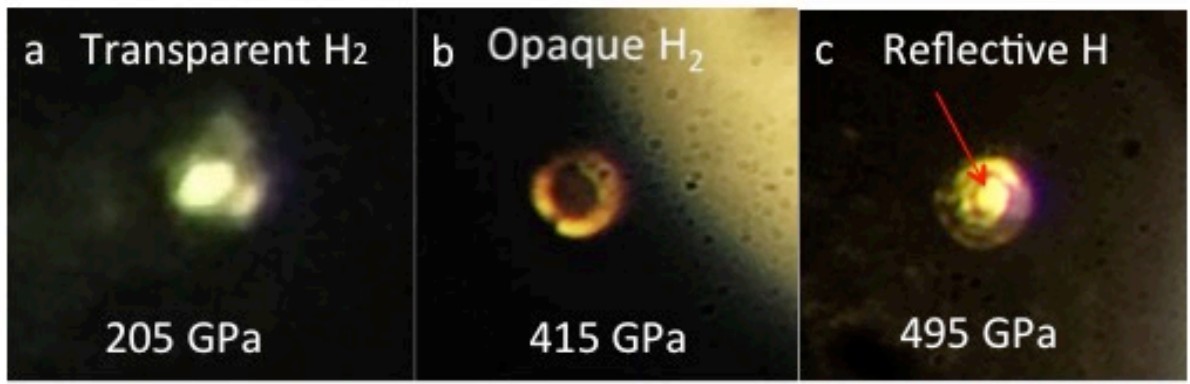

Fig. 2. Photographs of hydrogen at different stages of compression. Photos were taken with a smartphone camera at the ocular of a modified stereo microscope, using LED illumination in the other optical path of the microscope. a) At pressures to 335 GPa hydrogen was transparent. The sample was both front and back illuminated in this and in b; the less bright area around the sample is light reflected off of the Re gasket. b) At this stage of compression the sample was black and non-transmitting. The brighter area to the upper right corner is due to the LED illumination which was not focused on the sample for improved contrast; c) Photo of SMH at a pressure of 495 GPa. The sample is non-transmitting and observed in reflected light. The central region is clearly more reflective than the surrounding metallic rhenium gasket. The sample diameter is approximately 8 microns with thickness 1.2 microns (see SM).

At this point the only traditional way available for determining the pressure was to use Raman scattering of the diamond phonon line in the highly stressed culet region. For fear of diamond failure due to laser illumination and possible heating of the black sample, we decided to use the rotation of the screw as an indication of pressure. After some turns (see Fig. S3) the sample reflectance changed from black to high reflectivity, characteristic of a metal, shown in Fig. 2c. We then studied the wavelength dependence of the reflectance of the sample at liquid nitrogen and liquid helium temperatures, shown in Fig. 3. In order to do this the stereo not only allowed visual observation, but also laser illumination of the sample and formation of an image of the sample outside of the Macroscope (Fig. S5).

In our optical system (see SM) the sample is imaged onto a color CMOS camera (Thorlabs DC1645C). One can select the area of interest (this is effectively spatial filtering) and measure the reflectance from different surfaces. Thus, we measured the reflectance from SMH and the rhenium gasket. Reflectance was measured at three wavelengths in the visible spectral region, using both broadband white light, and three narrow band lasers that illuminated the sample (techniques are described in the SM). The measured reflectances are shown in Fig. 3, along with measurements of reflectance of the Re gasket, and reflectance from a sheet of Re at ambient conditions, that agreed well with values from the literature (*21*). At high pressure the stressed culet of the diamond becomes absorptive due to closing down of the diamond band gap (5.5 eV at ambient) (*22*). This attenuates both the incident and reflected light and is strongest in the blue. Fortunately this has been studied in detail by Vohra (*23*) who provides the optical density for both type I and II diamonds to very high pressures. We have used this (see Fig. S4) and show the corrected reflectance in Fig. 3. Finally, after measuring the reflectance, we used very low laser power (642.6 nm laser wavelength) and measured the Raman shift of the diamond phonon to be 2034 $cm^{-1}$. Using the linear scale of AK2006 (*17*) gives a pressure of 495±13 GPa when the sample was SMH (see Fig. S2, and discussion in SM).

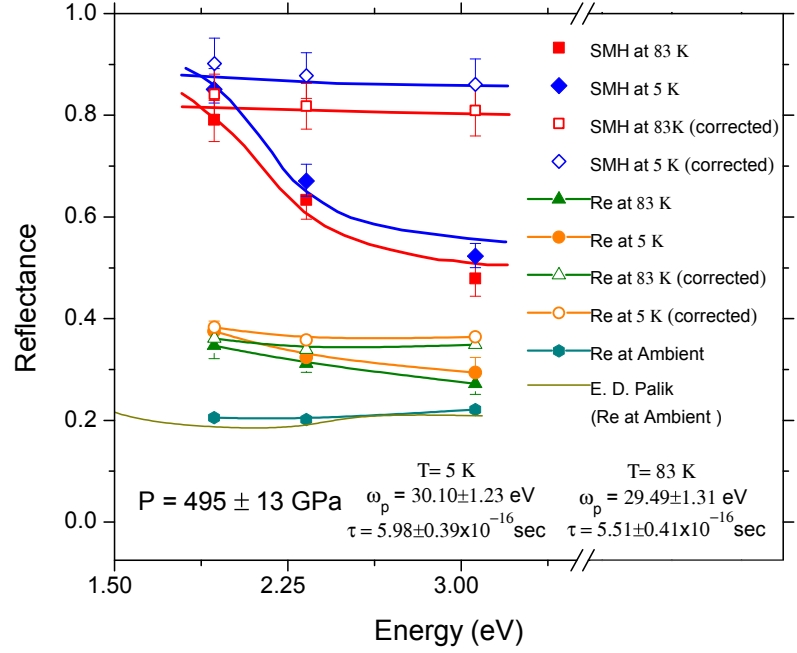

Fig. 3. The energy dependence of the normal incidence reflectance off of SMH and the rhenium gasket (P=495 GPa) at liquid nitrogen and liquid helium temperatures. We also show the reflectance from a surface of Re at a pressure of 1 bar at room temperature. The reflectances have been corrected for absorption in the diamond. Filled points are raw data and hollow ones are corrected. The uncertainties in the data points are from measurement of the reflectance and the correction procedure and represent random errors. The lines through the SMH data points are fits with a Drude free electron model; the lines through the Re data points are guides to the eye.

A very successful and easy to implement model is the Drude free electron model of a metal (*24*). The structure of SMH is predicted to be I4$_1$/amd (*5, 6*). A recent band structure analysis by Borinaga et al (*9*) shows that for this structure, electrons in SMH are close to the free-electron limit, which supports the application of a Drude model. An analysis of the reflectance using this model can yield important information concerning the fundamental properties of a metal.

The Drude model has two parameters, the plasma frequency $\omega_p$, and the relaxation time $\tau$. The plasma frequency is given by $\omega_p^2 = 4\pi n_e^2/m_e$ where $m_e$ and $e$ are the electron mass and charge, and $n_e$ is the electron density. The complex index of refraction of SMH is given by $N_H^2 = 1 - \omega_p^2/(\omega^2 + j\omega/\tau)$, where $\omega$ is the angular frequency of the light. SMH is in contact with the stressed diamond that has an index of refraction $N_D$; this has a value of ~2.41 in the red part of the spectrum at ambient conditions. We measured reflectance $R(\omega)=|(N_D-N_H)/(N_D+N_H)|^2$ as a function of energy or (angular) frequency $\omega$. A least-squares fit to the corrected reflectance data was used to determine the Drude parameters; at 5.5 K $\omega_p=30.10\pm1.23 eV$ and $\tau=5.98\pm0.39\times10^{-16}$ sec. These values differ significantly from a fit to the uncorrected data.

Since the diamond culet is stressed, the index of refraction in the region of contact with the SMH is expected to change from the value at ambient pressures and this might lead to a significant uncertainty in the fitting parameters. The index of the diamond under pressure and uniaxial stress has been studied by Surh et al (*25*). For hydrostatic pressures the pressure dependence is rather weak; however, for uniaxial stressed diamond the change of index can be substantial. We fitted data for values of 2.12 and 2.45, estimated from the work of Surh et al, for extreme stress. This resulted in an uncertainty in the Drude parameters that was much smaller than that due to the uncertainty in the measured reflectance (see Fig. 3). We fit the reflectance using a value $N_D=2.41$, yielding the values for the Drude parameters shown in Fig. 3 and Table I.

In Fig. 4 we show the P-T phase diagram of hydrogen along Pathway I. It has been speculated that the melting line (see Fig. 1) might extrapolate to the T=0 K limit at high pressure and metallic hydrogen might be a liquid in this limit. The vibron spectra in phase $H_2$-PRE correspond to spectra from a solid (*15*), and recent theory predicts a solid state (*5*). We believe that we have observed solid-not liquid- metallic hydrogen in the low temperature limit. Finally, in Fig. 4 we show a phase line for SMH based on the two points at T=83 and 5.5 K. Since the pressure was changed in larger increments by rotating a screw, there could be a possible systematic uncertainty of about 25 GPa on the low-pressure side of the phase line. See the SM for a further discussion on the pressure.

The plasma frequency $\omega_p=30.1\pm1.23$ is related to the electron density and yields a value of $n_e=6.7x10^{23} particles/cm^3$. We want to compare this to the atom density $n_a$ at a pressure of 495 GPa, however this is not known experimentally. There are a number of estimates of the density that range from ~6.6 to 8.8x10$^{23}$ particles/cm$^3$ (see SM). This is consistent with one electron per atom, so SMH is atomic metallic hydrogen, or the Wigner-

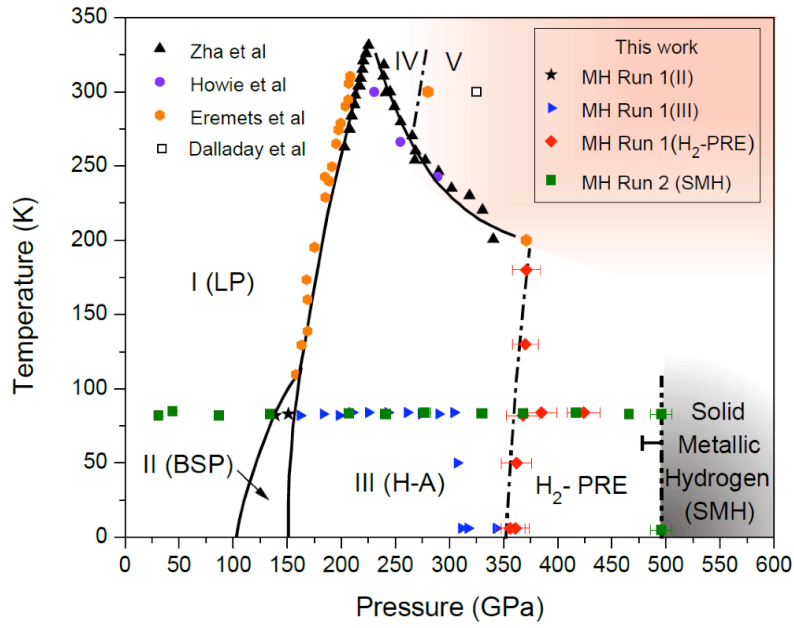

Fig. 4. The T-P phase diagram of hydrogen along Pathway I of Fig.1. The data (This work) shows the thermodynamic pathway that was followed for our measurements. We also show other recent data for te phases at lower pressures from Zha et al (*26*), Howie et al, Eremets et al (*16*), and Dias et al (*15*). A phase transition reported by Dalladay-Simpson et al (*27*) at 325 GPa is plotted as a point, however it has been shown that the data did not support the claim (*28*).

Huntington phase. SMH at 495 GPa is about 15-fold denser than zero-pressure hydrogen. In Table I we compare solid atomic hydrogen to other elements in the first column of the periodic table, and see a remarkable contrast in properties.

As of the writing of this article we are maintaining the first sample of the first element in the form of solid metallic hydrogen at liquid nitrogen temperature in a cryostat. This valuable sample may survive warming to room temperature and the DAC could be extracted from the cryostat for greatly enhanced observation and further study. Another possibility is to cool to liquid helium temperatures and slowly release the load to see if SMH is metastable. An important future measurement is to study this metal for high temperature superconductivity.

| ELEMENT | $n_e = \dfrac{m_e \omega_p^2}{4\pi e^2}$ ($10^{22}$/cm$^3$) | $n_a$ ($10^{22}$/cm$^3$) | $r_s/a_0$ | Plasma Frequency (eV) |
|---|---|---|---|---|
| H | 65.7 | 66.5-86.0 | 1.255-1.34 | 30.10 |
| Li | 3.68 | 4.63 | 3.25 | 7.12 |
| Na | 2.36 | 2.54 | 3.93 | 5.71 |
| K | 1.00 | 1.33 | 4.86 | 3.72 |

**Table I: Elements of the first column of the periodic table**. We compare the electron density (calculated from the plasma frequency) in the second column to the atom density in the third column, and see that there is about 1 electron/atom. The plasma frequencies are from Ref. (*29*). The data is for hydrogen at 5.5 K; all other elements are at 77 K. $r_s/a_0$ is defined in the SM.

We thank Mohamed Zaghoo, Ori Noked, Rachel Husband, and Chad Meyers for valuable assistance and discussions. Ron Walsworth provided a facility for annealing our diamonds. The NSF, grant DMR-1308641 and the DoE Stockpile Stewardship Academic Alliance Program, grant DE-NA0003346, supported this research. Preparation of diamond surfaces was performed in part at the Center for Nanoscale Systems (CNS), a member of the National Nanotechnology Infrastructure Network (NNIN), which is supported by the National Science Foundation under NSF award no. ECS-0335765. CNS is part of Harvard University. The data reported in this paper are tabulated and available from the authors. Both authors contributed equally to all aspects of this research.

# Observation of the Wigner-Huntington Transition to Solid Metallic Hydrogen

Ranga Dias and Isaac F. Silvera

## Materials and Methods

High purity hydrogen gas (four 9s) was cryogenically loaded into a diamond anvil cell in a cryostat similar to one described in Ref. (*1*), but with $CaF_2$ IR transmitting windows. An IR light beam from a Nicolet Fourier Transform infrared interferometer was focused onto the sample and then imaged onto an indium gallium arsenide detector for transmission studies in the near IR. Spectra (shown ahead) confirmed the sample as hydrogen and were used for pressure determination at intermediate pressures up to 335 GPa. The experimental optical table also incorporated standard instrumentation to measure ruby fluorescence.

**Pressure Determination**

There is no standard for the determination of the pressures produced in DACs, but there are a number of methods and calibrations that are used. The high-pressure community has adopted the ruby scale in which the frequency of the peak of the R1 fluorescence line is calibrated against pressure. Ruby is very useful for pressures up to 150-200 GPa, but is difficult to excite for higher pressures, although techniques exist (*3*). When improved calibrations are produced, older results can be recalibrated or scaled, if the frequencies or the calibration are given for a current measurement. Thus, pressures based on the current calibrations are used to characterize new phenomena.

Researchers studying hydrogen and its isotopes at extreme pressures have often used the Raman or IR peaks whose frequencies have been calibrated against pressure. This procedure was done in this study for pressures in the range 88 to 335 GPa. In Fig. S1 we show some of the IR vibron spectra that were used. In the region 135 to 335 GPa we used the pressure dependence of the $H_2$ IR-vibron measured by Zha et al (*4*) to determine the pressure of our sample. More detail can be found elsewhere (*2*).

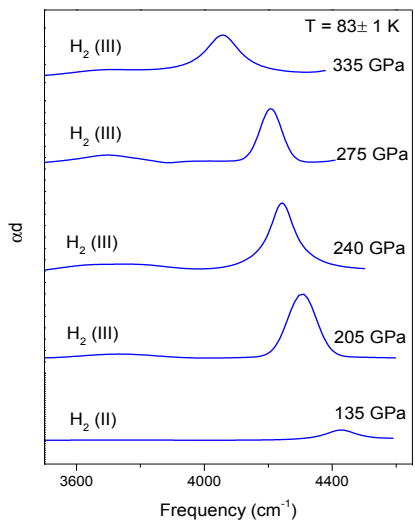

Fig. S1. IR vibron absorption peaks in solid hydrogen for several pressures. The phases are identified in parentheses.

For the highest pressures, researchers have started using the Raman scattered phonon spectra arising from the stressed region of the culet, proposed many years ago (*5, 6*); several calibrations exist. The highest pressure calibrations have been by Akahama and Kawamura (*7-9*). Due to the large uncertainties at higher pressures (*10, 11*) we round our pressure values to the nearest 0 or 5 GPa.

We have used the pressure scale of Akahama and Kawamura for the diamond phonon. Their latest calibrations are from the years 2006, 2007 and 2010, with highest measured pressures of 310, 370, and 410 GPa, respectively. The 2006 scale fits the pressure to a linear function of the frequency of the Raman feature, while the 2007 and 2010 fit to a quadratic equation, the same in both papers. We used the linear Raman scale to measure the pressure of SMH. The appropriate Raman feature (minimum of the derivative of the spectrum at the high frequency edge) was 2034 cm$^{-1}$, shown in Fig. S2. This yields P=495 GPa for the linear fit and 590 GPa for the quadratic fit. We decided to use the linear scale for reasons discussed ahead.

As explained in the text, for the highest pressures we used a pressure vs. load scale shown and explained in the caption of Fig. S3. The extrapolation of the diamond phonon to ~500-600 GPa is large; the data and our experience with our DAC using the pressure vs. load scale motivated us to use the more conservative linear scale.

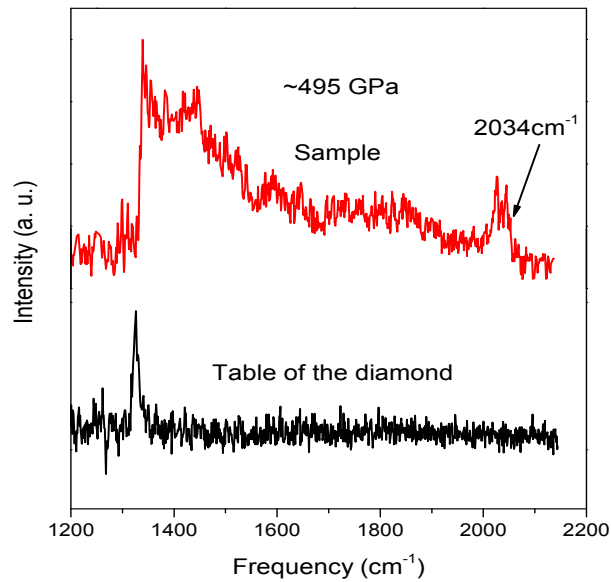

Fig. S2. The Raman shift of the SMH sample at the highest pressure, determined to be 495 GPa, as well as a spectrum from the unstressed table of the diamond.

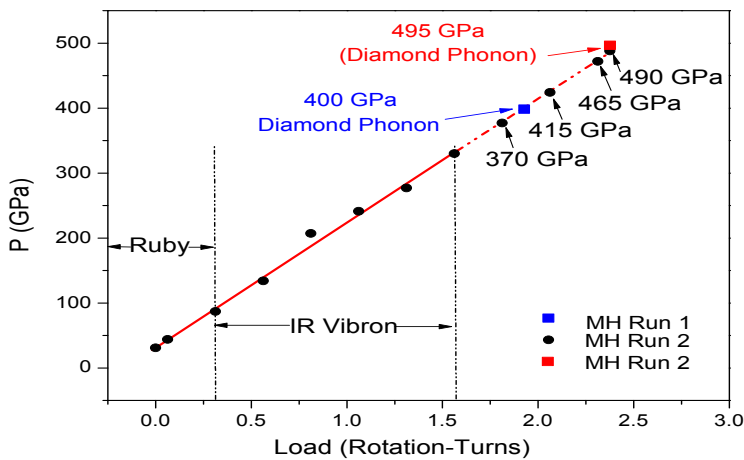

Fig. S3. Pressure as a function of load or rotation of the screw on the DAC. The solid round data points are measured pressures, confirming the rotation scale. The solid red line shows how this scale fits to a straight line, while the dash-dot part is an extrapolation of the load scale. The sample was visually examined at points shown on the extrapolated line. Our experience with this secondary scale is that the rotation scale continues linearly until the increase in pressure with rotation saturates (i.e., the pressure does not increase). The solid red square at 495 GPa is the pressure determined from the diamond Raman phonon, using the linear fit of Akahama and Kawamura (AK). If we extrapolate the quadratic scale of AK 2010, the pressure is 590 GPa. In Ref. (*2*) a maximum pressure of 420 GPa was achieved, using the 2010 AK scale. If we revert to the 2006 scale that pressure would be 400 GPa, shown by the blue square. Because our rotation scale and the 2006 scale show good agreement we have conservatively chosen to use AK2006 in this paper.

**Reflectance Corrections**

The unstressed diamond has a band gap of 5.5 eV. As pressure is increased the band gap closes and at very high pressures diamond becomes absorbing in the blue region, so the diamond window begins to close down (*12*). The behavior has been studied in detail in type I and II diamond to pressures as high as 421 GPa by Vohra (*13*). His article provides the optical density (OD) of the diamond by measuring reflectance of metals in a DAC. We measure exactly the same property, the reflectance of SMH. Thus, we can use the values of the two-pass OD (light in and out undergoes two attenuations) to correct our measured values of the reflectance. We measure reflectance at higher pressures than the OD scale of Vohra. Using his OD curves, we linearly extrapolate to our pressure for the frequencies used in the reflectance measurements. The extrapolation is shown in Fig. S4.

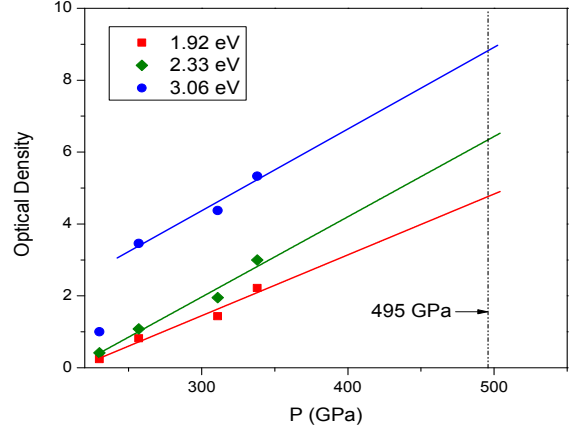

Fig. S4. The two-pass optical density of stressed diamond at high pressure used to correct the reflectance spectra for the three measured wavelengths. We have used a linear extrapolation of the data from Vohra.

**Particle Density Estimates of SMH**

There are many estimates of the density of SMH at the transition. We have used several of these, experimental and theoretical, to estimate the particle density of SMH at the pressure of the transition, which we take as 495 GPa. We give this as particle density and $r_s$, scaled to the Bohr radius $a_0$ ($1/n = (4\pi/3)(r_s a_o)^3$). Loubeyre et al measured the EOS to 109 GPa (*14*); extrapolating their fitting formula gives $8.1 \times 10^{23}$ atoms/cm$^3$ ($r_s$=1.25); extrapolation of the gap to zero in Ref. (*15*) gives $8.65 \times 10^{23}$ atoms/cm$^3$ ($r_s$=1.255); Ashcroft (*16*) calculated $6.65 \times 10^{23}$ atoms/cm$^3$ ($r_s$=1.34); McMahon and Ceperley (*17*) calculated $8.6 \times 10^{23}$ atoms/cm$^3$ ($r_s$=1.23). We summarize stating that the particle density at 495 GPa is in the range 6.65-8.6 $\times 10^{23}$ atoms/cm$^3$ ($r_s$=1.255-1.34).

**Optical Setup**

The optical setup used to measure the reflectance and Raman spectra of the diamond is shown in Fig. S5. Either the CMOS camera was used for reflectance measurements or a fiber optic coupled to the spectrometer was placed at the focus of the Wild Macroscope, along with an edge filter in the collimated beam emerging from the port of the Macroscope. For reflectance measurements two other lasers could be coupled in on the same beam path as the one shown, using flip-mirrors. The Macroscope also has a white light source that illuminates the sample at the focus. Not shown is the stereo microscope that could replace the Macroscope, used for visually monitoring the sample and photographing it.

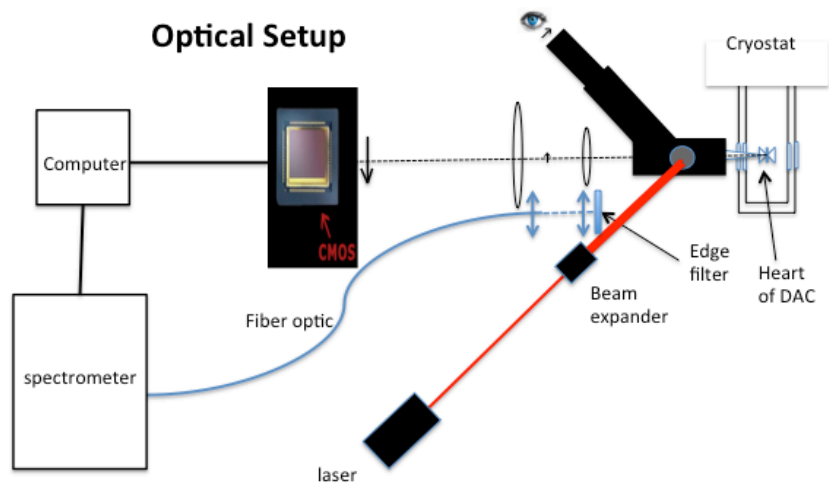

Fig. S5. The optical set up used to photograph and measure the reflectance of the sample in the DAC. There are two CaF$_2$ windows between the Macroscope and the DAC in the cryostat.

**Sample Dimensions**

We estimate the thickness of SMH as follows. The starting dimensions of the gasket hole were diam=24 µ and thickness=8 µ. The diameter of the gasket containing SMH was 8 µ, measured with the Macroscope. The final density was ~15 times greater than the zero pressure density. Assuming no loss of sample yields a final thickness of ~1.2 microns.

**Pathway II. Liquid Metallic Hydrogen.**

Recently liquid metallic hydrogen (LMH) has been produced at static pressures in a DAC, with a determination of the T-P phase line shown in Fig. 1 (*18*). Earlier, Weir et al (*19*) observed LMH in a shock experiment, however they saw a continuous change from insulating to metallic conductivity. Subsequently a number of shock experiments have been carried out, mainly on deuterium, observing metallic reflectance (*20-25*). A recent paper by McWilliams et al (*26*) disputes all of these findings of metallic behavior. Silvera et al (*27*) have found a flaw in the measurement technique of McWilliams et al, which likely resolves this conflicting observation.